# Reotomik Yüzeylerle Açıklık Geçen Alometri Bazlı Tasarım Sistemleri

Demircan Taş
İstanbul Teknik Üniversitesi
www.demircantas.com, demircantas@gmail.com

## Özet

Bu çalışmada mevcut yapılara dahil, halihazırda aralarında kolay geçiş imkanı bulunmayan iki veya daha fazla yatay düzlemi köprüleyecek tasarımların hızla üretilmesini mümkün kılma hedefiyle oluşturulan ikinci dereceden, malzeme tabanlı tasarım yöntemi, ve bu yöntemle üretilmiş üç örnek tasarım incelenecektir. Önerilen yöntemin ilk öğesi, kullanıcı tarafından, mevcut kullanım yüzeyleri üstünde tanımlanan eğrilerin ara değer hesabıyla yeni kullanım yüzeylerini, ve bu yeni yüzeyleri taşıyacak dolgu hacimlerini, reotomik (minimal) yüzeylerle oluşturan bir üretken sistemdir. İkinci öğe ise üretken sistemin oluşturduğu taşıyıcı hacme ait açık bırakılmış kalınlık değişkeninin, kullanıcı girdisi, bağlamsal veri, veya simülasyon sonuçları doğrultusunda, alometri ilhamlı bir yer değiştirme algoritmasıyla somutlaştırılmasıdır. Yöntemlerin oluşturulmasında, temel üretim metodu olan üç boyutlu yazım süreçlerine uygunluk gözetilmiştir. Eklemeli üretim yöntemlerinde kullanılan eriyik malzemenin katılaşma sürecinde geçici taşıyı olarak kullanılan destek geometrilerine olan ihtiyacı azaltma amacıyla, tasarım süreci de üretim ortamına benzer biçimde düşey olarak ilerletilmiştir. Önerilen sistem, simülasyon sonuçlarına uygun biçimde açık bırakılmış, iki boyutlu bir girdi ile taşıyıcı yapının birikim değişkenini besleyerek, yapının yüksek stres altındaki kısımlarda malzeme miktarı veya niteliğini değiştirerek dayanımı arttırmayı sağlar.

Yöntemlerin uygulanmasıyla farklı yatay düzlemleri birleştiren, üç fiziksel model elde edilmiştir. Modeller dijital ve fiziksel simülasyonlarla test edilebilir, sonuçlar tasarım sürecine geri katılarak her yinelemede sonucu geliştiren bir yapı oluşturabilir.

***Anahtar Kelimeler:*** *Alometri, Reotomik Yüzeyler, Üretken Sistemler, Malzeme Tabanlı, İkinci Dereceden Tasarım*

# Allometry Based Design Systems for Bridging via Rheotomic Surfaces

Demircan Taş
İstanbul Technical University
www.demircantas.com, demircantas@gmail.com

## Abstract

This study aims to present a material based, second order design method that makes the rapid creation of bridging structures in order to connect two or more horizontal planes which are currently separated and three design instances created via such method. The first element of the presented method is a generative system that creates circulation surfaces through the interpolation of the curves defined on the current surfaces, and also creates the structural volumes via rheotomic (minimal) surfaces. The second constituent is the instantiation of the exposed variable, connected to a displacement algorithm inspired by allometry based on user input, contextual data, or simulation results. The method was created with applicability through additive manufacturing in consideration. The design process – similar to manufacturing – proceeds in a vertical manner, in order to reduce the generation of support geometry as much as possible. The system includes a raster data input viable for simulation results, feeding the accumulation variable in order to modify material amount or quality with the aim of improving structural performance where stress is greater.

Through the application of the method, three physical models which connect different horizontal planes were obtained. The models can be evaluated with digital of physical simulations, and results can be utilized in an iterative manner, improving the results by each recursion.

## 1. Giriş

Mimari tasarım süreçlerinin bağlamsal ve gömülü veri ile geliştirilmesi, meslek içinde yaygınlaşan bir eğilimdir. Farklı seviyelerde veri toplanıp, tasarımı biçimlendirebilecek bilginin oluşturulmasında kullanılabilir. Saha ve malzeme bilgisine ek olarak yapı ve tasarım yöntembilimleri, tasarımı hesaplamalı yöntemlerle oluşturmaya yardımcı olabilir (Oxman, 2012).

## 2. Sorun Tanımı ve Amaç

Tasarım süreci, amaçların belirlenmesi, çeşitli çözümlerin üretilmesi, ve üretilen çözümlerin tanımlanan amacı tatmin edip etmediğinin değerlendirilmesi olarak tanımlanabilir. Sezgiler ve tümevarım yoluyla çözümler, gereklilikleri karşılayacak şekilde üretilebilir veya uyarlanabilir (Carrara, 1994).

Bu çalışma, yapı ve kent ölçeğinde farklı düzlemleri birbiri ile ilişkili hale getirecek, parametrik köprü/rampa yapıları oluşturabilen, uyarlanabilir, malzeme tabanlı (Oxman, 2012) bir üretken sistem ortaya koyma amacıyla gerçekleştirilmiştir.

## 3. Hedef

İkinci dereceden tasarım, mimarın tek bir tasarım ürünü üretmek yerine, belirli bir bağlamın farklı örneklerine (instance) uygun tasarımları ortaya koyacak bir sistemi ve kuralları kurgulamasını ön görür (Woods, 2001). Çok sayıda birimin benzer bağlamda, fakat farklı saha verilerine sahip durumlar altında üretilebilmesi hedefiyle çalışmada ikinci derece tasarım yaklaşımı benimsenmiştir.

## 4. Araçlar

Uyarlanır bir sistem, reotomik yüzeyler ile canlandırma tekniklerinin ve alometri ile birikme kavramlarının bir araya getirilmesiyle türetilmiştir.

Kullanıcı tarafından kenar veya yaylar biçiminde soyutlanmış başlangıç ve bitiş kesitleri ve hareket aksını tarifleyen vektörler, mevcut kenarlara gömülü eğriler kümesinde tanımlanır (Stiny, 2006). Sistem, kullanıcı tarafından tanımlanan başlangıç ve bitiş kesitleri, ve yön vektörleri üzerinden ara değer hesabıyla ara kesitleri oluşturur. Ara kesitlerin birleşimiyle kullanım yüzeyi elde edilir (Piker, 2009). Sonraki aşama için gerekli kavramsal kabuk değişken olarak tanımlanabilir, veya kullanıcı tarafından modellenebilir. Taşıyıcı sistem için önceden tanımlanmış, tekrarlı reotomik doku, kullanım yüzeyinin iz düşümü ve kavramsal kabuk ile boole kesişimi yoluyla işleme girdiğinde taşıyıcı sistem ön biçimi elde edilir. Ön tanımlı reotomik dokunun afin dönüşüm değerleri, değişkenler olarak tanımlıdır.

**Canlandırma, Reotomik Yüzeyler**

Eklemeli üretime ve parametrik yöntemlere uygunlukları sebebiyle reotomik (minimal) yüzeyler, kullanım yüzeyi ve hacim doldurma algoritmasını oluşturmakta tercih edilmiştir.

Ara değer hesabıyla elde edilen hareketin ve uzun pozlu fotoğraf tekniğine benzer bir sürecin (animation snapshot) farklı parametrelerine bağlı olarak, kullanım yüzeyi ve taşıyıcı dolgu biçimleri üreten bir sistem elde edilmiştir.

**Alometri / Birikme**

Alometri, canlıların farklı uzuvlarının, veya uzuvları oluşturan öğelerin işleve bağlı olarak değişik oranlarda büyümesini formüle bağlayan biyoloji kavramıdır (Gould, 1966). Çalışmada, taşıyıcı sistem kesitleri, birikim (displacement) algoritması ve algoritmayı yöneten ızgara (raster) verilerine bağlı olarak oluşum sürecinde uyumlu biçimde değişim gösterir. Izgara verisi, UV koordinat sistemleri ile bir arada kullanıldığında olası simülasyon ve analiz bulgularını yüksek verimle iletebilir. Bu sayede dijital yapı modeli simülasyon döngüleri ile beslenerek canlı kemik dokusuna benzer biçimde, zayıf kalan bölgeleri belirleyip birikme ile kuvvetlendirebilir.

## 5. Süreç

Çalışma, biçimlerin alometri yaklaşımı ile üretilmesini benimseyen bir yaklaşımla başlamıştır. Eş çap değerlerine sahip çemberler, ve bu çemberlerin belirli iç koordinatlarının (tek boyutlu U) G1 devamlılıkta eşlenmesi prensibine bağlı geçiş yüzeyleri tanımlanmıştır. Çemberlere zamana bağlı olarak yerel Z (ön) ekseninde afin büyüme, Y ekseninde (yukarı) afin dönme işlemi uygulanmıştır. Geçiş yüzeyleri parametrik olarak tanımlandığından animasyona, ek girdi gerekmeden uyum sağlayabilmektedir. Çemberler ve geçiş yüzeylerinin toplamından oluşan ana kesite Y düzleminde afin yer değiştirme işlemi, yine zamana bağlı olarak tanımlanmıştır.

Elde edilen hareketli kesitin zaman düzleminde eş aralıklarla dondurulması, ve bu kesitlerin ara değer hesabıyla kavramsal yüzeyler, bu yüzeylerin yerel koordinatlarında (U,V) tanımlanan kafese (grid) kalınlık verilmesiyle yapı modeli elde edilmiştir. **Şekil 1**'de değişken kesitin, nihai modelle ilişkisi gösterilmiştir.

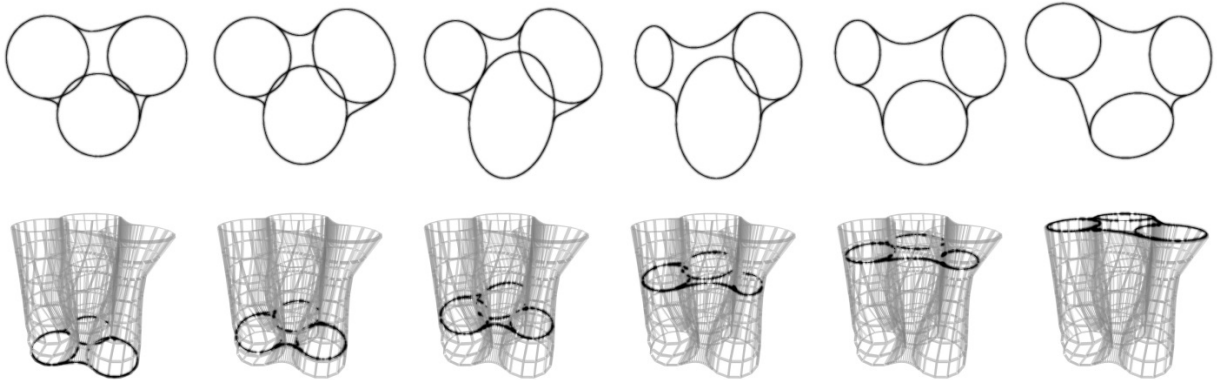

**Şekil 1:** Afin değişimleri uygulanan çember geometrileri ve geçiş eğrilerinden yapısal modelin oluşumu.

İlk aşamada elde edilen kabukların yapısal niteliğini geliştirmek amacıyla, üç boyutlu geometrilerin yerel UV koordinatlarında tanımlanabilen kafes (raster) nitelikte verilerden beslenen bir yer değiştirme (displacement) algoritması kullanılarak, birikim etkisi elde edilmiştir. Bu yöntem iki boyutta tanımlanabilen verilere bağlı olarak hacimlerin şişmesini sağlamaktadır. **Şekil 2**'de yöntem voronoi bazlı bir hacim modeli üzerinde denenmiştir. Aynı yöntemin ilk modele uygulanışı **Şekil 3**'de görülebilir.

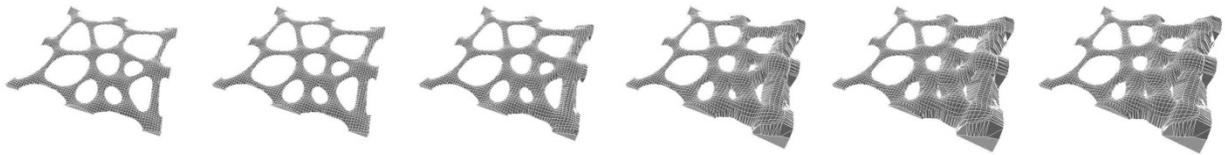

**Şekil 2:** Alometri bazlı birikim algoritmasının voronoi kafes üzerinde denenmesi.

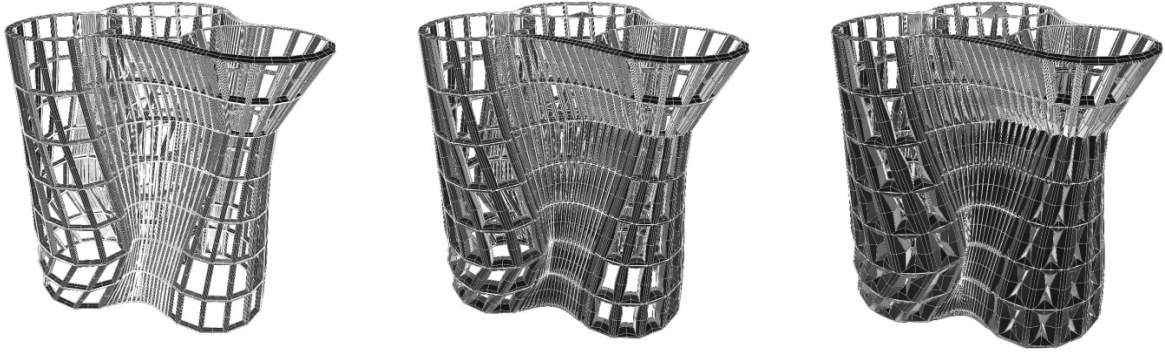

**Şekil 3:** Birikim algoritmasının kavramsal yapı üzerinde uygulanışı.

Çalışmanın bir sonraki aşamasında reotomik yüzeyler, ve bu yüzeylerin oluşum stratejileri araştırılmıştır. Bir doğru parçasının Y ekseninde afin yer değiştirme ve dönme sürecinin yüzeylenmesiyle spiral nitelikte temel yüzey elde edilmiştir (bkz. **Şekil 4)**.

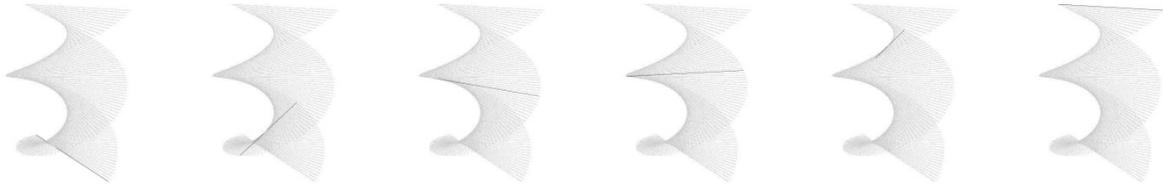

**Şekil 4:** Temel reotomik yüzeyin elde edilişi.

Elde edilen yüzeye temel bir kalınlık değeri verilerek helezonik bir hacim, bu hacimle tanımlı bir kare prizmasının boole kesişim işlemine sokulmasıyla tekrarlı birleşimlere uygun bir hacimsel yapı öğesi elde edilmiştir. Temel öğe, simetrik tekrarlar, afin dönüşümler ve istenilen kavramsal hacimlerle boole kesişimler uygulanarak uyumlu destek yapıları oluşturabilmektedir (bkz. **Şekil 5**).

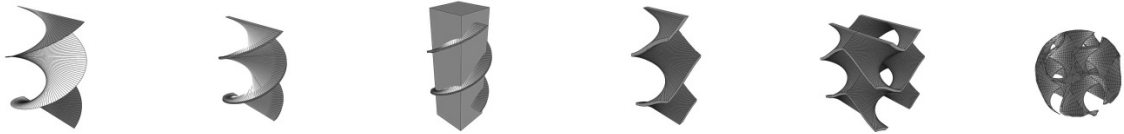

**Şekil 5:** Reotomik yüzeyin boole uygulamalarıyla tekrarlı ve biçimlenebilir taşıyıcı dokuya dönüşümü.

Temel reotomik yüzeye ek olarak, başlangıç aşamasında tanımlanan kenarların bükülmesi, ve bükme işlemi parametresinin geçilecek açıklığın U boyutunda değişken halde tanımlanması ile işleve bağlı olarak kesit değiştirebilen kullanım yüzeyleri tanımlanabilmektedir. Bu yüzeylerin taşınması için, kullanıcı tarafından kavramsal hacimler, iz düşüm ve boole kesişimleriyle tanımlanmaktadır. Tanımlanan hacimler uyumlu destek yapısıyla doldurulduktan sonra yapı dokusuna kullanıcı tarafından istenilen eksende afin dönüşümler uygulanabilmektedir.

Yapısal dokunun yerel UV koordinatları, geçilen açıklığın U koordinatına bağıl olarak düzlemsel projeksiyonla elde edilmektedir. Yapılan deneylerde bu koordinatların temas yüzeylerine uzaklığının tersi olarak tanımlanan parametrik bir raster veri bağlanmış, bu verinin birikim algoritmasını beslemesiyle kalınlık değiştiren yapısal hacim kesitleri elde edilmiştir (bkz. **Şekil 6,7,8**).

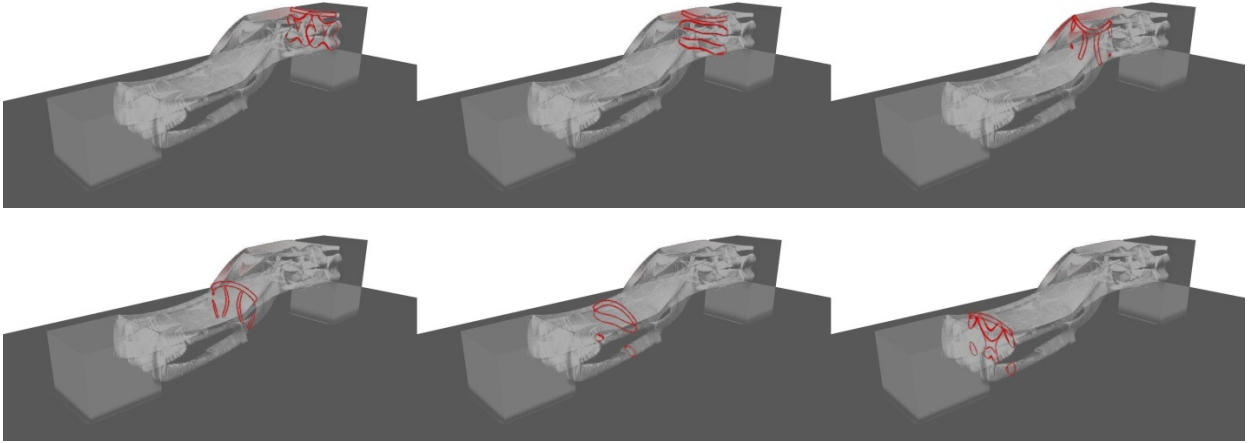

**Şekil 6:** Deney I'in değişken reotomik kesitleri.

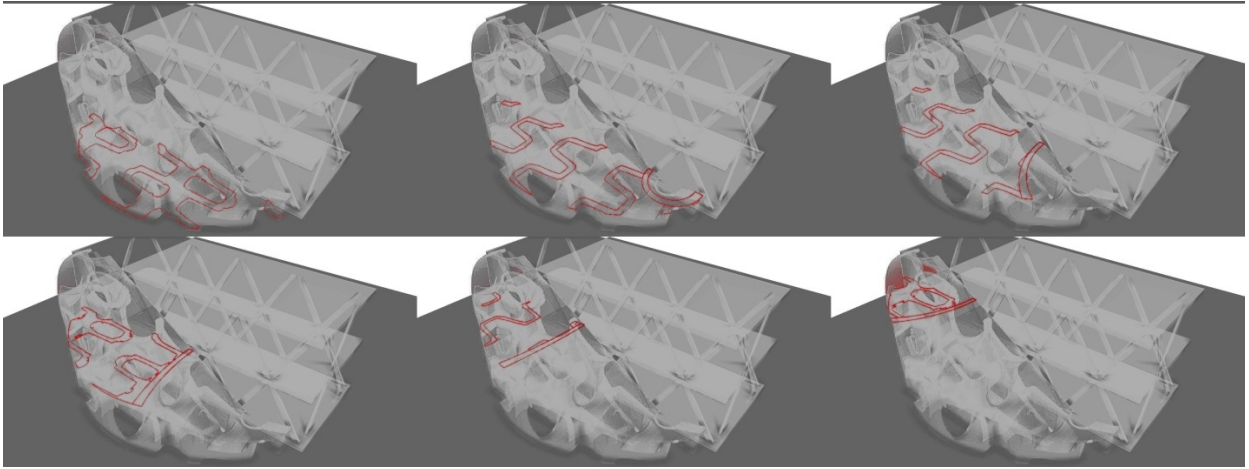

**Şekil 7:** Deney II'nin değişken reotomik kesitleri.

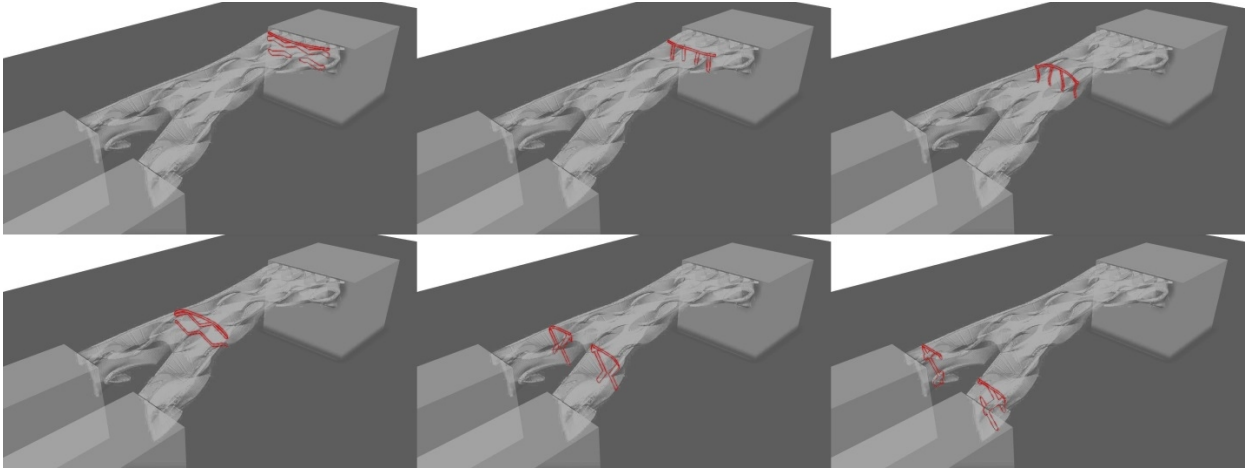

**Şekil 8:** Deney III'ün değişken reotomik kesitleri.

## 6. Bulgular ve Tartışma

Dijital modellerin taşıyıcı ve yüzey geometrilerinden, düşeyde katmanlama prensibiyle takım yolu verileri elde edilmiş, bu verilerin çift malzemeli bir üç boyutlu yazıcıda kullanımıyla fiziksel modeller oluşturulmuştur. Üç adet 1/100 ölçekli modele ek olarak birikim algoritmasını gözlemleyebilmek amacıyla bir adet 1/20 ölçekli model üretilmiştir. **Şekil 9** üretilen fiziksel modelleri göstermektedir.

Üretim sürecinde 1 ve 3 no'lu modeller neredeyse tümüyle destek yapısı gerektirmeden üretilebilmiş, 2 no'lu model ise yapısal dokunun yönelimi sebebiyle beklenenden yüksek miktarda destek malzemesi kullanımını gerektirmiştir. Dokunun yoğunluğu, destek malzemesinin temizlenmesini de zorlaştırmıştır. Bu zorluğun temel sebebi, y (düşey) ekseninde doğrusal olarak oluşturulan dokunu, spiral kabuk izdüşümüne indirgenmiş olmasıdır. İleri çalışmalarda taşıyıcı dokunun kabukla direkt uyum sergileyecek biçimler oluşturulmasına yönelik yöntemler incelenecektir.

Fiziksel modellerin kullanım nitelikleri incelendiğinde tüm modellerde değişken, 1 ve 3 no'lu modellerde doğrusal olmayan eğimler. Eğim, 1 no'lu modelin kenar akslarında, 2 no'lu modelin tümünde ve 3 no'lu modelin sınırlı bölgelerinde %50'ye yaklaşmaktadır. Modellerin üretim sürecinde yüzey eğimlerine yönelik bir değerlendirme kriteri mevcut olmadığından, elde edilen eğimler rastlantısal olmuştur. İleri çalışmalarda tanımlanacak performans kriterlerini denetleyecek sistemlerin kurgulanması amaçlanmaktadır.

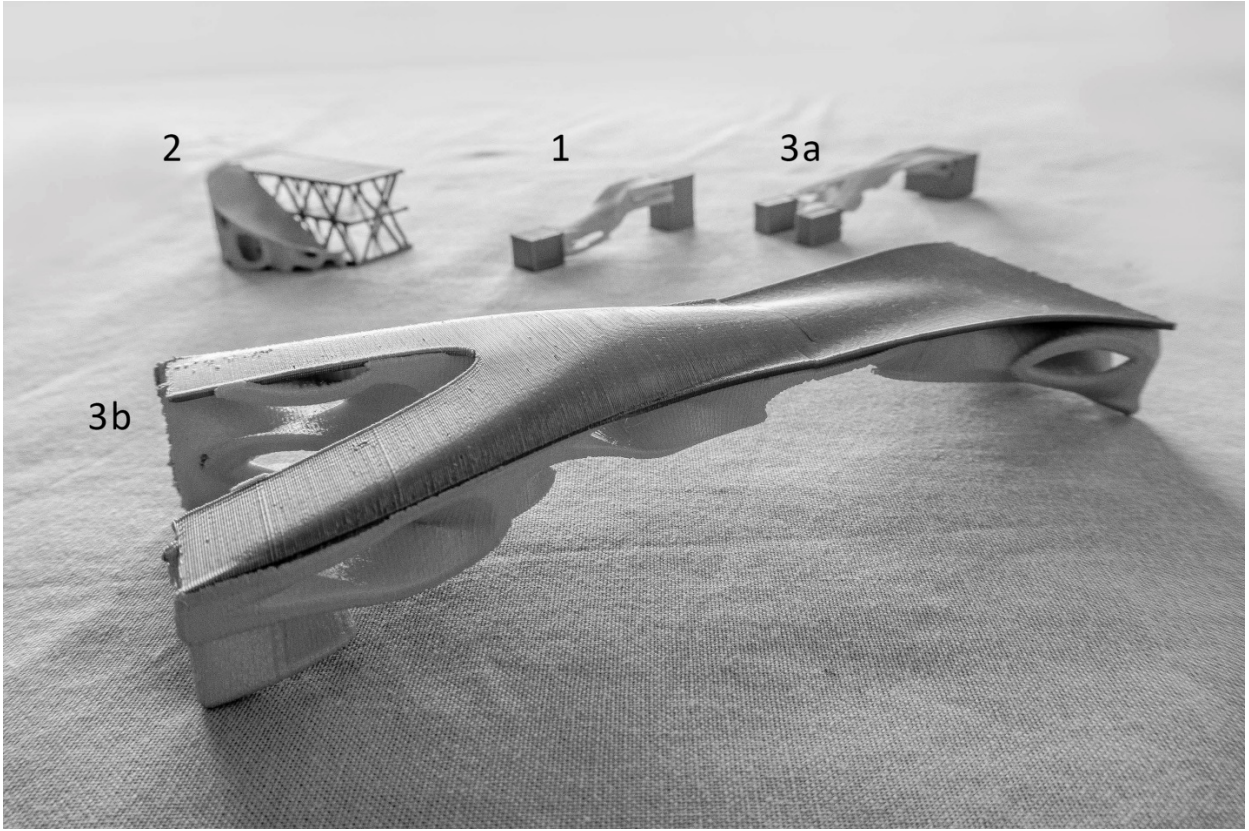

**Şekil 9:** Fiziksel modeller.

Üretim sonrasında 1/20 ölçekte üretilen model (3b) X ve Z eksenlerinde dönel strese maruz kaldığında hasar görmüştür. Oluşan çatlaklar üretimde benimsenen katmanlı takım yoluyla paralel niteliktedir. Yapısal dayanımla ilgili daha nitelikli veri toplamak amacıyla ileri çalışmalarda sınırlı eleman analizleri, ve / veya fiziksel modeller üzerinde kontrollü testler uygulanması hedeflenmektedir. Ek olarak, üç eksenli eklemeli üretim yöntemlerinin sebep olduğu paralel takım yollarının Z ekseninde dayanımı düşürdüğü gözlemlenmiştir. Üretim eksenlerinin değiştirilmesi ise, üretim kolaylığından taviz vermeyi gerektirmektedir. İleri çalışmalarda robotik üretim yöntemleri ile değişken eksenlerde tabakalaşma oluşturmak hedeflenmektedir.

Genel olarak üretken algoritmanın daha kısa zamanda yüksek tasarım permütasyonları oluşturacak şekilde optimize edilmesi, analiz ve simülasyonlarla, elde edilen tasarımların hedef / işlev uzayında konumlandırılması, ve tasarım döngüleriyle pareto nondominant eğriye yaklaşım hedeflenmektedir.

## 7. Sonuç

Reotomik yüzeyler ve alometrinin, döngüsel kullanımıyla bir üretken sistem oluşturulmuş, sistem farklı örneklerin hızla oluşturulmasında kullanılmıştır. Kullanım yüzeyi ve taşıyıcı doku eklemeli imalat yöntemleriyle prototiplenmiş, üretken sistemin fiziksel çıktıları incelenmiştir.

Hem dijital, hem fiziksel modellerin oluşum süreci hızlı ve ekonomik olmakla birlikte, benimsenen yöntem çok sayıda prototipin üretilmesini ve değerlendirilmesini gerektirmektedir. İlerleyen çalışmalarda özellikle dijital üretimlerin süresini daha da kısaltmak, değerlendirme kriterlerini uygulayacak yazılımları oluşturmak, ve robotik üretim tekniklerini araştırmak amaçlanmaktadır.

# KAYNAKLAR